\documentclass[fleqn,usenatbib]{mnras}

\usepackage[T1]{fontenc}

\DeclareRobustCommand{\VAN}[3]{#2}
\let\VANthebibliography\thebibliography
\def\thebibliography{\DeclareRobustCommand{\VAN}[3]{##3}\VANthebibliography}

\usepackage{graphicx}	% Including figure files
\usepackage{amsmath}	% Advanced maths commands
\usepackage{amssymb}	% Extra maths symbols
\usepackage{wasysym}           
\usepackage{graphicx}
\usepackage{epstopdf}
\usepackage{mathrsfs}
\usepackage{anyfontsize}
\usepackage{natbib}
\usepackage{color}
\usepackage{lipsum}
\usepackage{diagbox}
\usepackage{indentfirst}
\usepackage{newtxtext,newtxmath}

\title[On tidal capture for PNTs and QPEs]{Tidal capture of stars by supermassive black holes: implications for periodic nuclear transients and quasi-periodic eruptions}

\author[Cufari, Nixon, \& Coughlin]{
M.~Cufari$^{1}$\thanks{E-mail:mcufari@syr.edu},
C.~J.~Nixon$^{2,3}$,
and Eric~R.~Coughlin$^{1}$
\\
$^{1}$ Department of Physics, Syracuse University, Syracuse, NY 13244, USA\\
$^{2}$ School of Physics and Astronomy, University of Leicester, Leicester LE1 7RH, UK\\
$^{3}$ School of Physics and Astronomy, University of Leeds, Leeds LS2 9JT, UK
}

\date{Accepted XXX. Received YYY; in original form ZZZ}

\pubyear{2022}

\begin{document}
\label{firstpage}
\pagerange{\pageref{firstpage}--\pageref{lastpage}}
\maketitle

\begin{abstract}
Stars that plunge into the center of a galaxy are tidally perturbed by a supermassive black hole (SMBH), with closer encounters resulting in larger perturbations. Exciting these tides comes at the expense of the star's orbital energy, which leads to the naive conclusion that a smaller pericenter (i.e., a closer encounter between the star and SMBH) always yields a more tightly bound star to the SMBH. However, once the pericenter distance is small enough that the star is partially disrupted, morphological asymmetries in the mass lost by the star can yield an \emph{increase} in the orbital energy of the surviving core, resulting in its ejection -- not capture -- by the SMBH. Using smoothed-particle hydrodynamics simulations, we show that the combination of these two effects -- tidal excitation and asymmetric mass loss -- result in a maximum amount of energy lost through tides of $\sim 2.5\%$ of the binding energy of the star, which is significantly smaller than the theoretical maximum of the total stellar binding energy. This result implies that stars that are repeatedly partially disrupted by SMBHs many ($\gtrsim 10$) times on short-period orbits ($\lesssim$ few years), as has been invoked to explain the periodic nuclear transient ASASSN-14ko and quasi-periodic eruptions, must be bound to the SMBH through a mechanism other than tidal capture, such as a dynamical exchange (i.e., Hills capture).
\end{abstract}


\section{Introduction}
\label{sec:intro}
\noindent{}Binary stars may form when two field (i.e., individual) stars pass close to one another on nearly parabolic trajectories. As the stars pass, tidal oscillatory modes are excited in the stars at the expense of the stars' orbital energy. If the encounter is sufficiently close, the tides dissipate enough orbital energy to bind the stars to one another \citep{fabian75, press77}. 

A star may also be scattered into the region of parameter space, known as the ``loss-cone,'' that brings the star's distance of closest approach very near a supermassive black hole in the nucleus of a galaxy \citep[SMBH;][]{frank76,lightman77,cohn78}. Reasoning analogously to the binary formation scenario, one is tempted to conclude that stellar orbits with distances of closest approach nearer the black hole are more strongly tidally perturbed by, and thus more tightly bound to, the SMBH. However, there is a fundamental upper limit as to how efficient this process can be, because tides cannot transfer more energy into oscillations than the binding energy of the star itself. Additionally, as the pericenter distance of the star continues to decrease the perturbative tidal limit breaks down, and the star starts to lose a fraction of its outer envelope (known as a partial Tidal Disruption Event; TDE). Recent simulations have demonstrated that the asymmetric ejection of tidal debris in a partial TDE results in the \emph{unbinding} of the surviving stellar core from the SMBH \citep{manukian13,gafton15}. Therefore, the maximum binding energy a star can achieve through tidal interactions with an SMBH could be substantially smaller than the theoretical limit of the star's own binding energy. This maximum binding energy generated through tidal dissipation sets a fundamental timescale over which one would expect repeating partial TDEs -- a star that is bound to a SMBH that is partially destroyed on each pericenter passage (e.g., \citealt{zalamea2010, campana2015,miniutti2019,payne2021}) -- to recur if the bound star is supplied through tidal dissipation.

In this paper we present the results of smoothed particle hydrodynamics (SPH) simulations of partial TDEs from which we obtain the maximum binding energy. In Section \ref{sec:method} we describe the setup of the simulations, and in Section \ref{sec:results} we present the results. In Section \ref{sec:disc} we discuss the implications of our findings in the context of periodic nuclear transients (PNTs) and quasi-periodic eruptions (QPEs). We summarize and conclude in Section \ref{sec:conclusion}.

\section{Simulations}
\label{sec:simulations}
\subsection{Methodology}
\label{sec:method}

\noindent{}We performed simulations of partial TDEs with the SPH code {\sc phantom} \citep{price18}. The star was modeled as a polytrope with a $\gamma = 5/3$, adiabatic equation of state, with mass $M_{\star} = 1M_{\odot}$ and radius $R_{\star} = 1R_{\odot}$. In each simulation the star was injected from a distance of $12~r_t$, where $r_{\rm t} = R_{\star}\left(M_{\bullet}/M_{\star}\right)^{1/3}$, from the SMBH of mass $M_{\bullet} = 10^6M_{\odot}$ with the center of mass on a parabolic trajectory. The impact parameter $\beta\equiv r_{\rm{t}}/r_{\rm{p}}$, where $r_{\rm p}$ is the pericenter distance of the star, was varied between $\beta = 0.4$ and $\beta = 0.8$ in steps of $\Delta \beta = 0.02$. Additional details of the simulation setup can be found in \citet{coughlin2015}, and the algorithms for, e.g., self-gravity can be found in \citet{price18}.

We performed simulations with $1.25\times10^5$, $10^6$ particles and $8\times10^6$ particles, and found very little deviation in the outcome. All results presented here are from the $8\times10^6$ particle runs. For the purposes of determining the properties of the surviving star, we calculate averages using only those particles that have a density greater than 1\% of the maximum global density within the simulation, $\rho_{\rm max}$; we define this subset of particles as the core.  For example, the location of the core is calculated as the the average position of the subset of particles that satisfy this criterion, and the distance from the SMBH, $r_{\rm c}$, is determined from the Euclidean norm of the position. The analogous statement applies to the core velocity and speed, $v_{\rm c}$. Over the range of $\beta$ we simulated this criterion excludes the tidally stripped tails, and reducing the fraction to $0.1$\% does not impact the results because the tidal tails have a substantially lower density than $\sim 1$\% of $\rho_{\rm max}$. Similarly, increasing the fraction to $10$\% has a negligible effect on the results. However, values $>10\%$ can lead to significant noise in the results because in this case there are too few particles that contribute to the average.

We run each simulation until the core energy has settled to a constant value that we measure for our results. For disruptions with $\beta < 0.6$, this time is $\lesssim 1$ day after pericenter passage. Simulations with $\beta > 0.6$ are run up to $\sim 5$ days post-pericenter to ensure that the energy has converged to a constant value. Running our simulations to later times has no affect on our results.

\subsection{Results}
\label{sec:results}
\begin{figure*}
    \centering
    \includegraphics[width=0.49\textwidth]{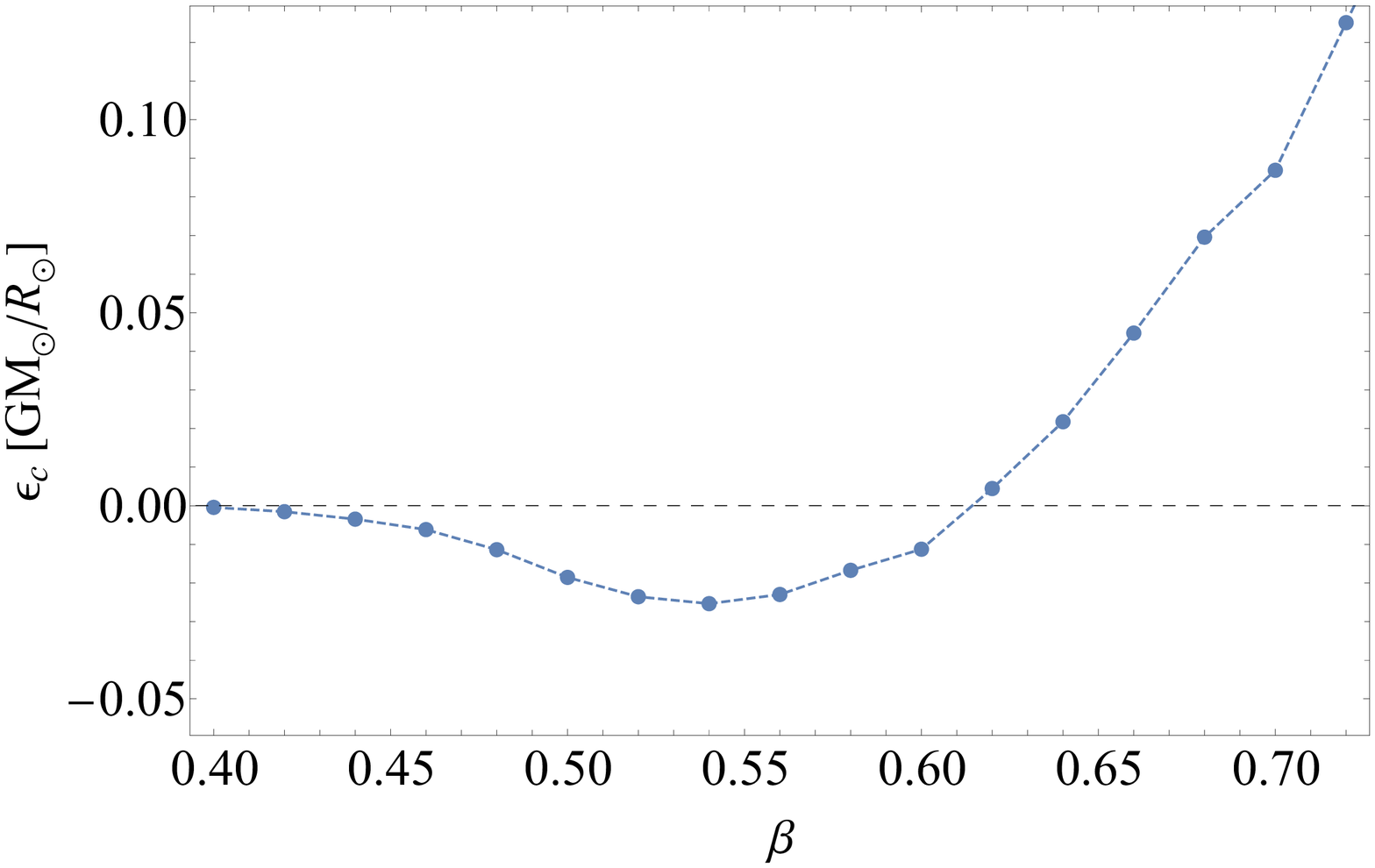}
      \includegraphics[width=0.49\textwidth]{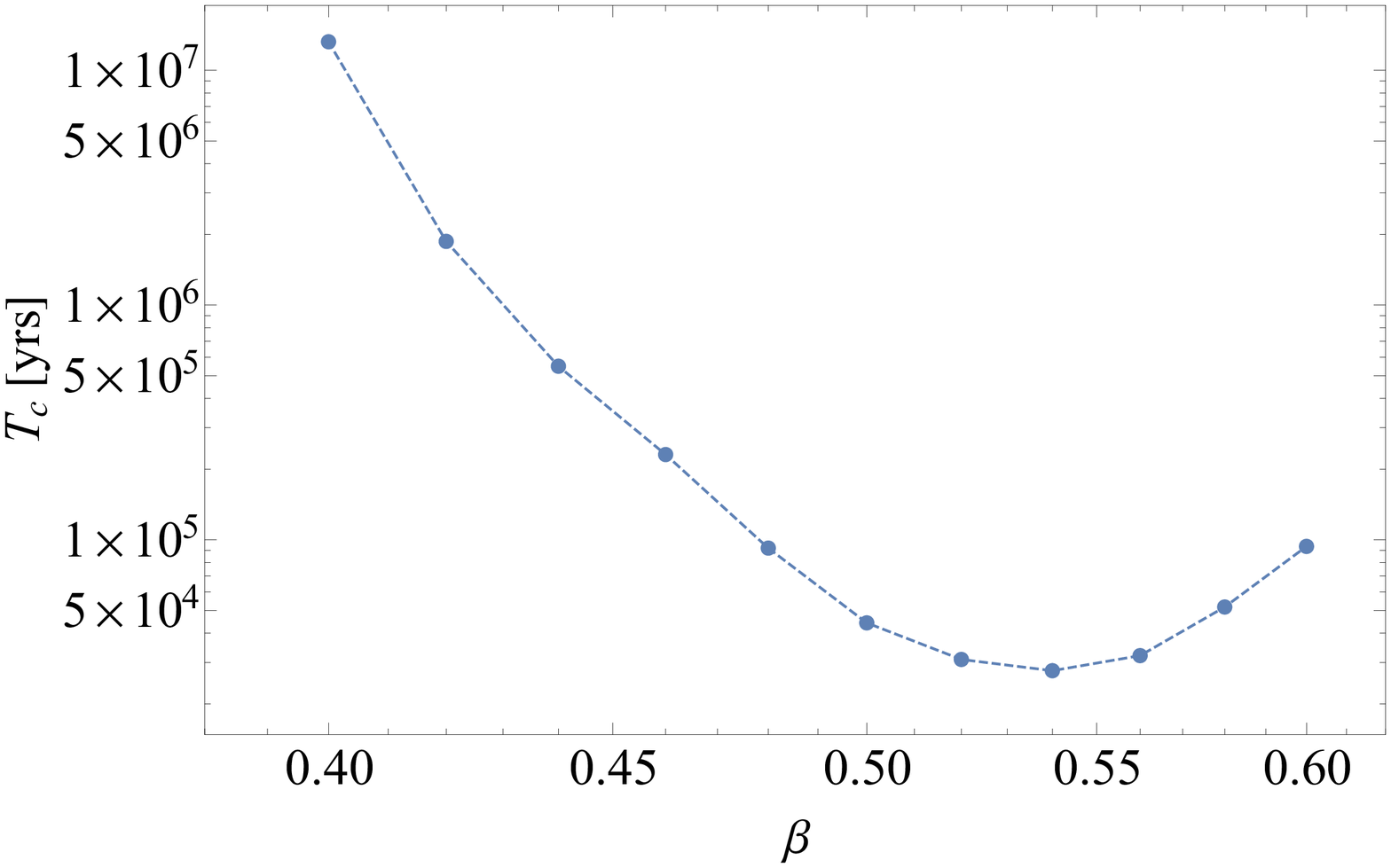}
    \caption{Left: The energy imparted to the star as a function of the impact parameter $\beta$. Stars with $\epsilon_{\rm c} < 0$ are bound to the black hole. There is a small range of $\beta$ for which mass loss occurs on first pericenter passage (i.e., $\beta > 0.5$) and for which $\epsilon_{\rm c} < 0$, the necessary condition for a repeating partial disruption to occur. Right: the period of the orbit of the star for encounters that yield a captured star as a function of the impact parameter $\beta$. The minimum period corresponds minimum orbital energy (i.e. the maximum binding energy), and for these parameters is $> 10^4$\,yrs.}
    \label{fig:etidal}
\end{figure*}

\noindent{}The left panel of Figure \ref{fig:etidal} shows the specific orbital energy of the core, calculated according to
\begin{equation}
    \epsilon_{\rm c} = \frac{1}{2}{v_{\rm{c}}^2} - \frac{GM_{\bullet}}{r_{\rm{c}}}.
\end{equation}
For $0.4 \lesssim \beta \lesssim 0.62$, the orbital energy is negative, implying that the surviving core is bound to the SMBH. The global minimum energy is $\sim 2-3\%$ of the binding energy of the star, i.e., 
\begin{equation}
    \epsilon_{\rm c, min} \simeq -0.025\frac{GM_{\star}}{R_{\star}}. \label{emin}
\end{equation}
The fact that the star becomes bound to the SMBH following its tidal interaction is in agreement with the classical calculations of tidal dissipation by \citet{fabian75, press77}. 

Conversely, all disruptions with $\beta \gtrsim 0.62$ result in a core with a net positive energy following the interaction with the SMBH, indicating that the star is on an unbound trajectory. This result agrees with those of \citet{manukian13, gafton15}, who found that the ``kick'' to the star could result in a velocity that is $\sim$ the escape speed of the star for $\beta$ very close to the value at which complete disruption occurs ($\beta \simeq 0.9$ for a 5/3 polytrope; \citealt{guillochon2013, mainetti2017}). Therefore, if a star reaches a pericenter distance with $0.62 \lesssim \beta \lesssim 0.9$, the star is not tidally captured but is instead tidally ejected, never to return to pericenter.

With the energy-semimajor axis and energy-period relationships for a Keplerian orbit, we find that the semimajor axis of the captured star (for $\beta \lesssim 0.62$, i.e., for stars that are not ejected) is
\begin{equation}
   a_{\rm c} = -\frac{GM_{\bullet}}{2\epsilon_{\rm c}} = \frac{1}{\eta_{\rm c}}\frac{R_{\star}}{2}\left(\frac{M_{\bullet}}{M_{\star}}\right), \label{ac}
\end{equation}
while the orbital period is
\begin{equation}
   T_{\rm c} = \frac{2\pi GM_{\bullet}}{\left(-2\epsilon_{\rm c}\right)^{3/2}} = \frac{1}{\eta_{\rm c}^{3/2}} \left(\frac{R_{\star}}{2}\right)^{3/2}\frac{2\pi}{\sqrt{GM_{\star}}}\left(\frac{M_{\bullet}}{M_{\star}}\right). \label{Tc}
\end{equation}
Here we defined the specific orbital energy of the captured star as $\epsilon_{\rm c} = -\eta_{\rm c}GM_{\star}/R_{\star}$, where $\eta_{\rm c} \lesssim 0.025$ (from Figure \ref{fig:etidal}). Using $M_{\bullet}/M_{\star} = 10^6$ in Equation \eqref{ac}, the minimum apocenter distance of the star (with $\eta_{\rm c} = 0.025$) is $\sim 1$ pc for $R_{\star} = 1R_{\odot}$.

The right panel of Figure \ref{fig:etidal} shows the orbital period of the captured star as a function of $\beta$ from the simulations and using Equation \eqref{Tc}. We see that the minimum orbital period of the captured star is $\sim 3\times 10^{4}$ yr. However, because of the strong dependence on $\eta_{\rm c}$ in Equation \eqref{Tc}, the period can be much larger for only small changes in $\beta$.

\section{discussion}\label{sec:disc}
\noindent{}Some current models for PNTs and QPEs suggest that they can be produced by stars on bound orbits about SMBHs with periods from hours to $\sim few$ years, with the star being partially disrupted and feeding an accretion flare every pericenter passage \citep{miniutti2019, king20, payne2021, king2022, wevers22}. PNTs have been suggested to have system parameters that are comparable to $M_{\bullet} =10^{7}M_{\odot}$, $R_{\star} = 1R_{\odot}$, $M_{\star} = 1M_{\odot}$, which gives, from Equation \eqref{Tc} with $\eta_{\rm c} = 0.025$,
\begin{equation}
    T_{\rm PNT} \simeq 3\times 10^{5}\textrm{ yr}, \label{TPNT}
\end{equation}
which is obviously too long to explain the observed periods $\lesssim 1$ year \citep{payne2021, wevers22}. Even if one could inject the theoretical maximum energy into the star, and thus set $\eta_{\rm c} = 1$ in Equation \eqref{Tc}, we would obtain $T_{\rm PNT} \simeq 10^{3}$ yr. As argued by \citet{cufari22}, even in this overly optimistic scenario, PNTs need an additional mechanism to bind the star sufficiently tightly to the SMBH to produce their observed periods in situ.

On the other hand, it is possible that a PNT could \emph{initially} be generated with an orbital period given by Equation \eqref{TPNT}, but the period then decays through gravitational-wave emission (over millions of years) to produce a period as short as $\sim 1$ year by the time that we observe it\footnote{As pointed out in \citet{payne2021}, the rate of change of the orbital period due to gravitational waves is actually too small to explain the observed value for ASASSN-14ko, but it should be possible for gravitational waves to shrink the orbit in general and for other systems.}. However, a problem with this interpretation is that even for the maximum value of $\eta_{\rm c} = 0.025$, Equation \eqref{ac} for the semimajor axis of the orbit yields an apocenter distance of $\sim 2a_{\rm c} \sim 10$ pc for $M_{\bullet}/M_{\star} = 10^{7}$. Thus, we would expect the star to interact with, and be once again perturbed by, the nuclear star cluster, and it seems very unlikely that the star will repeatedly return to the same pericenter over multiple passages. 

Additionally, while we have assumed that the star is on a parabolic orbit, it could have a velocity at infinity that is $v_{\infty} \simeq \sigma$, with $\sigma$ the galactic velocity dispersion \citep{miller05}. From Figure \ref{fig:etidal}, the maximum amount of energy able to be dissipated through tides is
\begin{equation}
\epsilon_{\rm c, max} \simeq 0.025\frac{GM_{\star}}{R_{\star}} \simeq \frac{1}{2}\left(\frac{M_{\star}}{M_{\odot}}\right)\left(\frac{R_{\star}}{R_{\odot}}\right)^{-1}\left(100\textrm{ km s}^{-1}\right)^2.
\end{equation}
For a $10^7M_{\odot}$ SMBH, the velocity dispersion from the $M-\sigma$ relation is $\sigma \sim 110$ km s$^{-1}$\citep{marsden20}. Thus, tides may not actually be capable of dissipating the true (positive) energy of the orbit, and hence binding the star through this mechanism may be impossible in the first place.

QPEs -- with periods of the order hours -- are typically modeled with a $\sim 10^{5}M_{\odot}$ SMBH partially disrupting a white dwarf star. For such a system, Equation \eqref{Tc} with $M_{\star} = 0.6 M_{\odot}$, $R_{\star} = 0.011R_{\odot}$ \citep{nauenberg72}, $M_{\bullet} = 4\times 10^{5}M_{\odot}$, and $\eta_{\rm c} = 0.025$ gives
\begin{equation}
    T_{\rm QPE} \simeq 28\textrm{ yr}, \label{TQPE}
\end{equation}
which shows that it is not possible to bind a star to a SMBH through tidal dissipation and immediately reproduce the observed orbital periods of QPEs. As for PNTs, it is nonetheless possible that gravitational-wave emission shrinks the orbit\footnote{The orbital period may also be reduced through the interaction with a pre-existing AGN disc \citep{syer91, cufari22, lu22}.} to $\sim$ hours before we detect them. Because the binding energy of a white dwarf is substantially larger than that of a main sequence star, the minimum apocenter distance is correspondingly smaller (from Equation \ref{ac}), and the maximum imparted energy through tides is substantially larger than the energy at infinity. Thus, without further investigation that is outside the scope of the present work, we cannot conclusively state whether or not an additional mechanism is required for producing the observed orbital periods in QPEs (under the paradigm that they are powered by repeatedly partially disrupted white dwarfs). 

On the other hand, the tidal breakup of a binary star system (i.e., Hills capture; \citealt{hills75}) can bind the star with a substantially shorter period (compared to just tidal dissipation) if the binary is sufficiently tight. As demonstrated by \citet{cufari22}, this is a plausible explanation for the origin of the $\sim 114$ day period of ASASSN-14ko. An additional argument against the gravitational-wave inspiral scenario for explaining ASASSN-14ko is that the mass lost by the star must be $\sim 1$\% of the mass of the star to power the emission \citep{cufari22}, and hence it cannot have survived many ($\gtrsim 100$ s) interactions prior to the $\sim 10$ that have been observed since the initial detection. Similarly, if the white dwarf binary separation is not much larger than the radius of the white dwarf itself, then Equation (5) from \citet{cufari22} with $a_{\star} = 0.04R_{\odot}$, $M_{\bullet} = 4\times 10^5M_{\odot}$, and $M_{\star} = 0.6 M_{\odot}$ yields an orbital period of $\sim 8.3$ hours for the period of the bound star following a Hills-capture event. As also suggested in \citet{cufari22}, it therefore seems plausible that the periods of QPEs can be generated with a dynamical exchange process without the need for additional dissipation through other means. 

\section{Summary and Conclusions}
\label{sec:conclusion}
\noindent{}We presented SPH simulations of the interaction between a star and an SMBH to determine the maximum degree by which a star can be bound to an SMBH through tidal dissipation. Two competing mechanisms prevent the star from becoming arbitrarily bound to the SMBH. As the distance of closest approach between the star and SMBH shrinks, energy from the star's orbit is expended in exciting dynamical tides in the star. However, once the star reaches a distance of closest approach comparable to its tidal radius, the star is kicked to positive energies as a result of asymmetry in the tidal tails that are liberated from the star. The competition between these two physical effects results in a minimum possible orbital energy of the star following the tidal encounter. For the parameters we simulated here, the location of this minimum is at $\beta \sim 0.55$ (pericenter distance of $r_{\rm t}/0.55$ with $r_{\rm t}$ the usual tidal radius) and the binding energy of the orbit is $\sim 2.5\%$ of the star's binding energy; see Equation \eqref{emin} specifically. This minimum energy is significantly smaller than the theoretical maximum, being the entirety of the stellar binding energy.

In order to produce a repeating partial tidal disruption via a tidal dissipation mechanism, the energy kick imparted to the star must be negative, otherwise the star will be ejected on a hyperbolic trajectory. Hence, it would appear that there is a relatively small region of parameter space within which the star is only partially destroyed, not ejected, and survives for many ($\gtrsim 10$) pericenter passages, specifically $0.4\lesssim \beta\lesssim 0.5$ for the type of star considered here. Initially non-rotating stars in this range of $\beta$ will be rotating at a nontrivial fraction of breakup following the initial interaction, which will move the effective tidal radius out \citep{golightly19}, but provided that the pericenter (effectively unaltered because of the small ratio of the maximal angular momentum of the star to the angular momentum of the orbit itself) is still within this tidal radius, the star may transfer a small amount of mass during each additional pericenter passage. In this manner, a star may undergo many cycles of partial disruptions before being destroyed or ejected.

In our simulations we modeled the star as a 5/3 polytrope, which is most applicable to low-mass main sequence stars and and low-mass white dwarfs. By number, most stars are thought to fall into this regime. However, more massive (radiative) stars are considerably more centrally concentrated than predicted by a 5/3-polytropic model. Here we have shown that the minimum energy for the captured star -- modeled as a 5/3 polytrope -- occurs at $\beta \approx 0.55$. This result will depend somewhat on the type of star being considered. For example, Figure 3 of \cite{manukian13} shows that it occurs for $\beta < 1$ when the star is modeled as a 4/3-polytrope and that there is very little dependence of the result on the black hole mass. \citet{faber05} considered the tidal capture of a planet by a star in which the mass ratio was $q = 0.001$, and found a minimum binding energy of $\sim 14\%$ of the binding energy of the planet at $\beta = 10/19 \simeq 0.523$ (see their Table 1). \citet{kremer22} also recently considered black hole-star systems with mass ratios closer to unity, and found a similar effect to the one described here if the mass ratio was $0.02$ or $0.05$ if the star was modeled as a $\gamma = 5/3$ polytrope, but that the star was able to go from bound to completely disrupted -- without being ejected -- as the mass ratio increased beyond 0.05 and the star was modeled with the Eddington standard model (see their Figure 1). We defer an analysis of the minimum orbital energy -- and the $\beta$ at which the minimum energy occurs -- as a function of the mass ratio and the type of star to future work.

In the context of extreme mass-ratio inspirals, \cite{zalamea2010} demonstrated that a white dwarf (or other compact object) completes thousands of large-eccentricity orbits before reaching the direct capture radius of the SMBH. However, their model accounts only for the orbital decay due to gravitational wave emission and omits tidal dissipation and mass loss asymmetry effects. Likewise, our simulations omit the effects of orbital decay due to gravitational wave emission. The pericenter distance of our simulated disruptions is $> 50~r_{\rm{g}}$, so orbital decay due to general relativistic effects (at least on the first pericenter passage) is negligible. For more compact stars with smaller tidal radii nearer the event horizon, orbital decay due to general relativistic effects will be more significant over fewer orbital periods (though the change in the pericenter will still be extremely small, as all of the dissipation occurs near pericenter for these high-eccentricity systems; thus the tidal interaction itself may be relatively unaltered, aside from the stronger tidal field of the SMBH due to relativistic gravity). Future work on partial TDEs nearer the horizon of the SMBH should incorporate the change in orbital energy due to tidal interaction and mass loss asymmetry alongside gravitational wave emission. 

Finally, here we focused on orbits that produce partial TDEs in the traditional sense, i.e., the tidal force is not sufficiently strong to destroy the star completely. However, \cite{nixon22} found that at very high $\beta$ (in their case $\beta = 16$), the compression experienced by the star near pericenter could revive self-gravity to the point that a core \emph{reformed} with a binding energy (to the SMBH) that was much larger than the value predicted by Equation \eqref{emin} (though we caution that while the mass contained in the core was converged, the orbital period -- and thus the binding energy -- was resolution-dependent in their simulations). While encounters with high-$\beta$ are rare\footnote{For example, Equation 16 of \cite{coughlin22b} shows that the fraction of TDEs with $\beta > 10$ for a $10^6 M_{\odot}$ Schwarzschild SMBH -- including general relativistic effects -- is 0.0046; note that this is a factor of $\sim 4$ smaller than the value derived by ignoring general relativistic effects, given by their Equation 17.}, it may be possible for tidal capture in this considerably more exotic scenario to produce shorter-period orbits than through the traditional means. 

\section*{Data Availability Statement}
\noindent{}Code to reproduce the results in this paper is available upon reasonable request to the corresponding author.
\section*{acknowledgements}
\noindent{}M.C.~acknowledges support from the Syracuse Office of Undergraduate Research and Creative Engagement (SOURCE). CJN acknowledges support from the Science and Technology Facilities Council (grant no. ST/W000857/1), and the Leverhulme Trust (grant no. RPG-2021-380). E.R.C.~acknowledges support from the National Science Foundation through grant no. AST-2006684 and the Oakridge Associated Universities through a Ralph E.~Powe Junior Faculty Enhancement Award. This work was performed using the DiRAC Data Intensive service at Leicester, operated by the University of Leicester IT Services, which forms part of the STFC DiRAC HPC Facility (www.dirac.ac.uk). The equipment was funded by BEIS capital funding via STFC capital grants ST/K000373/1 and ST/R002363/1 and STFC DiRAC Operations grant ST/R001014/1. DiRAC is part of the National e-Infrastructure.

\bibliographystyle{mnras}
\bibliography{refs}

\bsp	% typesetting comment
\label{lastpage}
\end{document}